\documentclass[preprint,12pt]{elsarticle}
\usepackage{CJK,graphicx}
\usepackage{mathrsfs}
\usepackage{graphicx}

\usepackage{amssymb}

\journal{{\rm $^*$Corresponding author. Email: yangli@iie.ac.cn}}

\begin{document}
\begin{CJK*}{GBK}{song}
\begin{frontmatter}



\title{Quantum probabilistic encryption scheme based on conjugate coding}

\author{Li Yang$^*$}
\author{Chong Xiang}
\author{Bao Li}

\address{State Key Laboratory of Information Security, Institute of Information Engineering, CAS, Beijing 100093, China}

\begin{abstract}
We present a quantum probabilistic encryption algorithm for a private-key encryption scheme based on conjugate coding of the qubit string. A probabilistic encryption algorithm is generally adopted in public-key encryption protocols. Here we consider the way it increases the unicity distance of both classical and quantum private-key encryption schemes. The security of quantum probabilistic private-key encryption schemes against two kinds of attacks is analyzed. By using the no-signalling postulate, we show that the scheme can resist attack to the key. The scheme's security against plaintext attack is also investigated by considering the information-theoretic indistinguishability of the encryption scheme. Finally, we make a conjecture regarding Breidbart's attack.
\end{abstract}




\end{frontmatter}


\newtheorem{theorem}{Theorem}
\newtheorem{lemma}[theorem]{Lemma}
\newtheorem{conjecture}[theorem]{Conjecture}
\newtheorem{corollary}[theorem]{corollary}
\newtheorem{definition}{Definition}
\newtheorem{proposition}[theorem]{Proposition}

\section{Introduction}
Public-key cryptosystems were first proposed in the 1970s \cite{Dif76,Riv78}. Because the original public-key encryption schemes were not secure under chosen-plaintext attack (IND-CPA), Goldwasser and Micali introduced the idea of probabilistic encryption\cite{God84} in 1984. Up till now, both quantum public-key encryption \cite{Yan03,Nik08,Gao08,Yan1012,Gotte} and deterministic quantum private-key encryption \cite{Boy03,Amb00,Amb04} have been investigated. In this paper, we first present a classical private-key encryption scheme with a probabilistic algorithm, then design a quantum probabilistic algorithm for a private-key encryption scheme. We shall show that the probabilistic algorithm can increase the unicity distance of both classical and quantum encryption schemes\cite{Wie83}.
Breidbart's attack on the four-state quantum cryp-tography scheme has been discussed for many years now. Bennett et al. \cite{Ben92} proved that Breidbart's attack is weaker than regular basis eavesdropping; Huttner and Ekirt \cite{Hut94} showed that it is more effective after performing a standard error correction; Yang, Wu, and Liu \cite{Yan02} improved the second result with an extended BB84 QKD protocol. These results are based on the effective average Alice/Eve mutual information. In this paper, we show that the POVM of Breidbart's attack yields the maximum classical trace distance. This implies that Breidbart's attack is the best measurement for the quantum bit string.

\section{The quantum probabilistic private-key encryption scheme}
\subsection{Classical private-key encryption with probabilistic algorithm}
Let the triplet $(\mathcal{E}, S, \mathcal{D})$ be a classical private-key encryption scheme, where $\mathcal{E}, \mathcal{D}$ are two polynomial-time classical algorithms, and $S$ is a set of classical keys. We construct the classical probabilistic private-key encryption scheme as a quintet $(\mathcal{H}, \Lambda, \mathcal{E}, S, \mathcal{D})$, where $\mathcal{H}_\lambda\in\mathcal{H}$ are polynomial-time algorithms indexed by $\lambda\in\Lambda=\{1,2\cdots,l\}$. Each $\mathcal{H}_\lambda$ has a polynomial-time inverse algorithm. Considering a block cipher with length $k$, which is the length of the key, let the plaintext $m$ be divided into $n$ blocks, $m=(m_1, m_2\cdots m_n$). The process then works as follows:
\begin{enumerate}
  \item The sender Alice and receiver Bob preshare a classical key $s\in S$.
  \item For the plaintext $m_i$, Alice first randomly chooses $\lambda_i$,and performs $\mathcal{H}_{\lambda_i}$ on this block, obtains $\mathcal{H}_{\lambda_i}(m_i)$.
  \item Alice performs $\mathcal{E}$ on $\mathcal{H}_{\lambda_i}(m_i)$ and $s$, and it outputs $C_{\lambda_i}=\mathcal{E}(\mathcal{H}_{\lambda_i}(m_i))$. Alice send $C_{\lambda_i}$ to Bob.
  \item Bob performs $\mathcal{D}$ on $C_{\lambda_i}$ and $s$, and obtains $H_{\lambda_i}(m_i)$. He then tries to decrypt $l$ times with different $\mathcal{H}_{\lambda'_i}^{-1}$ and gets $\mathcal{H}_{\lambda'_i}^{-1}(\mathcal{H}_{\lambda_i}(m_i))$, $\lambda'_i=1, \cdots, l$.
\end{enumerate}

Since there is more than one plaintext left after decryption, Bob has to select one making use of the redundancy of plaintext. A similar type scheme of public-key encryption had been considered by Rabin \cite{Rab79}.

For the encryption algorithm defined by $(\mathcal{E}, S, \mathcal{D})$, assume the time complexities $t_1$ for $\mathcal{E}$, $t_2$ for $\mathcal{D}$, $t_3$ for $\mathcal{H}_\lambda$, and $t_4$ for $\mathcal{H}_\lambda^{-1}$
\begin{enumerate}
  \item The new complexity for encryption is $n\times(t_1+t_3)$.
  \item The new complexity for decryption is $n\times(t_2+\frac{1}{2}l\times t_4)$.
  \item The new complexity for exhaustive attack is $2^k\times n\times (t_2+l^n\times t_4)$.
\end{enumerate}
This result shows that raising polynomial time complexity in encryption and decryption leads to exponentially growing time complexity for exhaustive attack.

\subsection{Quantum probabilistic private-key encryption scheme}
Assume the two parties Alice and Bob share a bit-string $s=s_1,s_2\cdots,s_k$ as their private key. Define 
$$\Omega_0^k=\{r\in\{0,1\}^k|r_1\oplus r_2\oplus\cdots\oplus r_k=0\},$$ and 
$$\Omega_1^k=\{r\in\{0,1\}^k|r_1\oplus r_2\oplus\cdots\oplus r_k=1\}.$$ 
The plaintexts are $m=m_1\cdots m_n$. Alice encrypts them bit by bit as follow:\\

{\bf [Encryption $\mathcal{E}$]}

\begin{enumerate}
\item Alice randomly selects $r^{(i)}\in\Omega_{m_i}^k$.
\item Alice prepares the quantum state
\begin{eqnarray}
    |\phi_{m_i}\rangle=|r_1^{(i)}\rangle_{s_1}\otimes\cdots\otimes|r_k^{(i)}\rangle_{s_k}.
\end{eqnarray}
\item Alice sends the state $|\phi_{m_i}\rangle$ to Bob.
\end{enumerate}

The density operator of the ciphertext encrypted from $m_i$ is:
\begin{eqnarray}
    \rho(m_i,s,r^{(i)})=|r_1^{(i)}\rangle_{s_1}\langle r_1^{(i)}|\otimes\cdots\otimes|r_k^{(i)}\rangle_{s_k}\langle r_k^{(i)}|.
\end{eqnarray}
where
\begin{eqnarray}\label{sta}
|0\rangle_0\equiv|0\rangle, |1\rangle_0\equiv|1\rangle, |0\rangle_1\equiv|+\rangle,|1\rangle_1\equiv|-\rangle.
\end{eqnarray}

After receiving the ciphertexts, Bob measures them using the private key. We can see that the state Bob gets is:
\begin{eqnarray}
    \rho_B=\frac{1}{N_B}\bigotimes_{i=1}^n\sum_{r^{(i)},m_i}\rho(m_i,s,r^{(i)}),
\end{eqnarray}
where $N_B=2^{n\times k}.$\\

{\bf [Decryption $\mathcal{D}$]}

\begin{enumerate}
\item Bob measures the ciphertext state using the value of $s$. The ciphertext state of $m_i$ will collapse to $r^{(i)}$ with probability 1.
\item Bob calculates $m_i=r_1^{(i)}\oplus \cdots\oplus r_k^{(i)}.$
\end{enumerate}

In this scheme, the attacker Eve does not have $s$, $r^{(i)}$, and $m$, so if she intercepts the quantum channel and gets the state, she will get the state:
\begin{eqnarray}
    \rho_E=\frac{1}{N_E}\bigotimes_{i=1}^n\sum_{s,r^{(i)},m_i}\rho(m_i,s,r^{(i)}),
\end{eqnarray}
where $N_E=2^{2k\times n}.$

\section{Attack to the key}
If we assume that the plaintext is completely random, we obtain the following:
\begin{lemma} If Eve had any method $\mathcal{F}$ for accessing the information in the key from the ciphertext state, This would contradict the no-signalling postulate.
\end{lemma}
{\bf Proof. }Let $|r\rangle_s=|r_1\rangle_{s_1}\cdots|r_k\rangle_{s_k}$ be the state obtained by Eve. This satisfy Eq. (\ref{sta}). Because the plaintext is truly random, so the $r_i$ must also be truly random.

Assuming Eve had a method $\mathcal{F}$ that could gain information about the key from the state of ciphertext, so that  $\mathcal{F}(|r\rangle_s)=s$, then Alice and Eve could achieve superluminal signalling with the entangled channel.
For $k=1$ the superluminal signalling process would be as follow:\\

{\bf [Superluminal signalling]}
\begin{enumerate}
\item Alice prepares entangled state $|\phi\rangle=\frac{\sqrt{2}}{2}(|00\rangle+|11\rangle)$, sends one party of it to Eve and keeps the other one herself.
\item If $b=0$, when Alice wants to transmit the information $b$, she uses the basis $|0\rangle,|1\rangle$ to measure the qubit she kept. Then $|\phi\rangle$ will collapse to $|00\rangle$ or $|11\rangle$ with probability 1/2 for each, and at the same time, the qubit Eve gets will collapse to $|0\rangle_0$ or $|1\rangle_0$ with probability 1/2 for each.
\item If $b=1$, the basis used to measure the qubit is $|+\rangle,|-\rangle$. Then $|\phi\rangle$ will collapse to $|++\rangle$ or $|--\rangle$ with probability 1/2 for each, so the qubit Eve gets will collapse to $|0\rangle_1$ or $|1\rangle_1$ with probability 1/2 for each.
\item Eve uses $\mathcal{F}$ to get the basis of the qubit, with the results $\mathcal{F}(|r\rangle_s)=s=b$.
\end{enumerate}

Hence Alice and Eve achieve superluminal signalling. If Alice wants to transmit more bits, she can share more entangled states and repeat this scheme time after time. The random collapse corresponds to the random distribution of the plaintexts.

As superluminal signalling is not allowed by the no-signalling postulate, there can be no such $\mathcal{F}$. $\Box$

When the plaintexts have a random distribution, direct attack on the key can be used to build a superluminal signalling scheme. In fact, the plaintexts cannot have this property. In this context, we thus make the following conjecture:

\begin{conjecture}If the plaintexts $m$ have a pseudo-random distribution, direct attack on the key $s$ should not be possible..
\end{conjecture}

This conjecture implies that, in order to make the scheme secure, we can transform the plaintext by a one-way trapdoor permutation such as RSA. On the basis of this conjecture, we get the following corollary:
\begin{corollary}
The scheme can resist the attack to the key if the plaintext $m$ has pseudo-random distribution.
\end{corollary}

\section{Attack to the plaintext}
If Eve attacks the information in the key $s$ with method $\mathcal{F}$, she can of course get the information in the plaintexts $m$. But even if Eve has a method $\mathcal{G}$ for accessing the information in $m$, this does not mean that she can get the information in $s$. We will show that Eve also has no such $\mathcal{G}$.

\subsection{Indistinguishability}
Goldrich defined indistinguishability for the classical private-key encryption scheme\cite{Gold04}. Here we define information-theoretic indistinguishability for a quantum private-key encryption scheme.
With definition 5.2.3 in \cite{Gold04} Goldrich defined the indistinguishability for private-key encryption of classical message.Here we propose the information-theoretic indistinguishability for quantum private-key encryption.

\begin{definition}\label{def}
A quantum encryption $(G,E,D)$ is information theoretically indistinguishable if for every quantum circuit family \{$C_n$\}, every positive polynomial $p(\cdot)$, all sufficiently large $n$'s, and every $x, y\in\{0,1\}^{poly(n)}$(i.e.,$|x|=|y|$),
\begin{eqnarray}\label{ITS}
\Big|\textrm{Pr}[C_n(E_{G(1^n)}(x))=1]-\textrm{Pr}[C_n(E_{G(1^n)}(y))=1]\Big|<\frac{1}{p(n)},
\end{eqnarray}
where the encryption algorithm $E$ should be quantum algorithm, $G$ is a internal coin tosser of algorithm, and the ciphertext $E(x), E(y)$ are quantum states.
\end{definition}
{\bf Remark 1.} In Subsection 5.5.2 of Ref.\cite{Gold04}, Goldrich states that his definition of indistinguishability for the classical private-key encryption scheme is computational when the classical circuit family is polynomial-size, and information-theoretic when the classical circuit family was no limits on size.

In both Ref.\cite{Yan10} and this paper, information-theoretic security or indistinguishability are all defined using a quantum circuit family \{$C_n$\} without size limits. Furthermore, we think of indistinguishability as a kind of security that can be classified by three different conditions:
\begin{enumerate}
  \item If the quantum circuit family \{$C_n$\} is polynomial-size, the above definition defines computational indistinguishability.
  \item If the quantum circuit family \{$C_n$\} has no size limits, it delivers the above definition.
  \item if the quantum circuit family \{$C_n$\} has a specifical exponential-size determined by the protocol, we define it as physical indistinguishability of protocol.
\end{enumerate}

This classification can also be extended to semantic security and non-malleability. It should be noted that the physical security here concerns the protocol, it is different from the physical security of the system, which means physical isolation of the security system. For example, the quantum bit commitment protocol in \cite{Yan10} is a physically secure scheme, because the unitary matrix for the attack operation is physically incomputable. In fact, physical security of algorithms can satisfy all the security requirements of human beings.

\begin{figure}[!htbp]
\center
  \includegraphics[width=8cm]{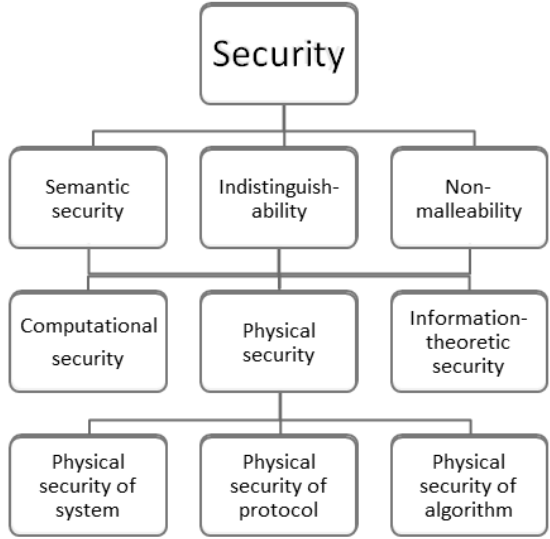}\\
  \caption{the classification of security}\label{cos}
\end{figure}

In fact, physical security of algorithms can satisfy all the security requirements of human beings.$\Box$

Next, we give a sufficient condition for information theoretic indistinguishability.

\begin{theorem}\label{the}For all plaintexts $x$ and $y$, let the density operators of cipher states $E(x)$ and $E(y)$ be $\rho_x$ and $\rho_y$, respectively. A quantum private-key encryption is said to be information theoretically indistinguishable if, for every positive polynomial $p(\cdot)$ and every sufficiently large $n$'s,
\begin{eqnarray}
D(\rho_x,\rho_y)<\frac{1}{p(n)}.
\end{eqnarray}
\end{theorem}

{\bf Proof.} Here we follow the proof of Theorem 1 in Ref. \cite{Yan10}). Define $S_x$ as the set of all states Eve could receive when the plaintext is x. For every quantum circuit family $\{C_n\}$,
\begin{eqnarray}
& &\textrm{Pr}[C_n(E_{G(1^n)}(x))=1]\nonumber\\
&= &\sum_{\rho_x^i\in S_x}p_i\cdot\textrm{Pr}[C_n(\rho_x^i\otimes \sigma)=1]\nonumber\\
&= &\textrm{Pr}[C_n(\sum_{\rho_x^i\in S_x}p_i\rho_x^i\otimes \sigma)=1]\nonumber\\
&= &\textrm{Pr}[C_n(\rho_x\otimes \sigma)=1],
\end{eqnarray}
where $\sigma$ is the density operator of service bits of $C_n$.

Similarly,
\begin{eqnarray}
\textrm{Pr}[C_n(E_{G(1^n)}(y))=1]=\textrm{Pr}[C_n(\rho_y\otimes \sigma)=1].
\end{eqnarray}

Every method of attack for distinguishing two density operators corresponds to a positive-operator-valued measure (POVM) \{$E_m$\}. Let $p_m=\textrm{Tr}(C_n(\rho_x\otimes \sigma)E_m)$, and $q_m=\textrm{Tr}(C_n(\rho_y\otimes \sigma)E_m)$ be the probabilities of measurement results labeled by $m$. Then we have:
\begin{eqnarray}
& &\Big|\textrm{Pr}[C_n(\rho_x\otimes \sigma)=1]-\textrm{Pr}[C_n(\rho_y\otimes \sigma)=1]\Big|\nonumber\\
&\leq & \max_{\{E_m\}}\frac{1}{2}\sum_m|\textrm{Tr}[E_m(C_n(\rho_x\otimes \sigma)-C_n(\rho_y\otimes \sigma))]|\nonumber\\
&= & \max_{\{E_m\}}D(p_m, q_m)\nonumber\\
&=&D(C_n(\rho_x\otimes \sigma),C_n(\rho_y\otimes \sigma)).
\end{eqnarray}

Since
\begin{eqnarray}
D(C_n(\rho_x\otimes \sigma),C_n(\rho_y\otimes \sigma))\leq D(\rho_x\otimes \sigma,\rho_y\otimes \sigma)=D(\rho_x, \rho_y)<\frac{1}{p(n)}, \end{eqnarray}
it follows that
\begin{eqnarray}
\Big|\textrm{Pr}[C_n(E_{G(1^n)}(\rho_x))=1]-
\textrm{Pr}[C_n(E_{G(1^n)}(\rho_y))=1]\Big|<\frac{1}{p(n)},
\end{eqnarray}
which, according to the Definition \ref{def}, proves the theorem. $\Box$

\subsection{Analysis of the scheme}
In the scheme, $r$ is a random string selected by Alice, so $r$ and the private key $s$ are unknown to Eve. Let $\rho_b=\rho_b^k$ be the density operator of cipher $\mathcal{E}(b)$, while the length of $s$ and $r$ is $k$. For Eve, the density operator $\rho_0^k$ should take all possible values of $s$ and $r\in\Omega_0^k$, and similarly, $\rho_1^k$ should take all possible values of $s$ and $r\in\Omega_1^k$. These density operators can be written as:
$$\rho_0=\frac{1}{2^{2k-1}}\sum_{r\in\Omega_0^k,s}|r_1\rangle_{s_1}\langle r_1|\otimes\cdots\otimes|r_k\rangle_{s_k}\langle r_k|,$$
and
\begin{eqnarray}\label{rho}
\rho_1=\frac{1}{2^{2k-1}}\sum_{r\in\Omega_1^k,s}|r_1\rangle_{s_1}\langle r_1|\otimes\cdots\otimes|r_k\rangle_{s_i}\langle r_k|.
\end{eqnarray}

It can be seem that
\begin{eqnarray}
\rho_0^k=\frac{\rho_0^{k-1}\otimes\rho_0^1+\rho_1^{k-1}\otimes\rho_1^1}{2}
\end{eqnarray}
\begin{eqnarray}\label{rho2}
\rho_1^k=\frac{\rho_0^{k-1}\otimes\rho_1^1+\rho_1^{k-1}\otimes\rho_0^1}{2}.
\end{eqnarray}

Then we have
\begin{eqnarray}\label{eq2}
& &D(\rho_0^k,\rho_1^k)\nonumber\\
&=&\frac{1}{2}tr|\frac{\rho_0^{k-1}\otimes\rho_0^1+\rho_1^{k-1}\otimes\rho_1^1}{2}-\frac{\rho_0^{k-1}\otimes\rho_1^1+\rho_1^{k-1}\otimes\rho_0^1}{2}|\nonumber\\
&=&\frac{1}{4}tr|(\rho_0^{k-1}-\rho_1^{k-1})\otimes(\rho_0^1-\rho_1^1)|.
\end{eqnarray}

For every density operator $\tau_1,\tau_2$, we have
$$|\tau_1\otimes\tau_2|=|\tau_1|\otimes|\tau_2|.$$

Then Eq.(\ref{eq2}) is equivalent to
\begin{eqnarray}
& &\frac{1}{4}tr(|\rho_0^{k-1}-\rho_1^{k-1}|\otimes|\rho_0^1-\rho_1^1|)\nonumber\\
&=&\frac{1}{2}tr|\rho_0^1-\rho_1^1|\times\frac{1}{2}tr|\rho_0^{k-1}-\rho_1^{k-1}|\nonumber\\
&=&D(\rho_0^1,\rho_1^1)\times D(\rho_0^{k-1},\rho_1^{k-1}).
\end{eqnarray}

Repeating the above process, we have
\begin{eqnarray}
D(\rho_0^k,\rho_1^k)=(D(\rho_0^1,\rho_1^1))^k.
\end{eqnarray}

Note that $$\rho_0^0=\frac{1}{2}(|0\rangle\langle0|+|+\rangle\langle+|)=
\frac{1}{4}\left[\begin{array}{cc}3 &1\\1 &1\end{array}\right],$$
$$\rho_1^0=\frac{1}{2}(|1\rangle\langle1|+|-\rangle\langle-|)=
\frac{1}{4}\left[\begin{array}{cc}1 &-1\\-1 &3\end{array}\right],$$

so we have $D(\rho_0^1,\rho_1^1)=\frac{\sqrt{2}}{2}$, and hence
\begin{eqnarray}\label{dis}
D(\rho_0^k,\rho_1^k)=(\frac{\sqrt{2}}{2})^k.
\end{eqnarray}

This trace distance refers to the encryption of one bit. As the security of the scheme is based on the trace distance between two plaintext states encrypted from any bit strings $x$ and $y$, there must be an upper bound $n_k$ determined by the length of the key $s$ such that, when the length of plaintexts $x$ and $y$ is less than $n_k$, the scheme can satisfy
$$D(\rho_x, \rho_y)\leq\frac{1}{p(n)}.$$

This bound, which makes the scheme secure, has not work out yet.

\section{Quantum unicity distance}
Even if the plaintexts have been coded with a one-way trapdoor permutation which makes them pseudo-random, they should still have redundancy. So when the ciphertexts encrypted by the same key are long enough, the scheme must be attacked by trying every possible key, which means our scheme has a limited unicity distance\cite{sha49}.

Assume that attacker Eve can distinguish a pseudo-random string when it is longer than a sufficiently large $N$. On this premise, if Eve gets more than $k\times 2^k\times N$ ciphertext states, she can divided these states into $2^k$ groups, each corresponding to more than $N$ plaintexts. Eve can then decrypt these groups of ciphertext states with $2^k$ different bit-strings taking from key space respectively. While the unique key yields a pseudo-random string, other keys all yield a random string. Finally, Eve will distinguish the decrypted string and access the private key.

Besides $N$, the quantum unicity distance of our scheme is at least an exponential function of $k$. This result is based on the improvement due to probabilistic encryption.

On the other hand, for a quantum deterministic private-key encryption scheme, the unicity distance should be much less than that of our scheme. We adopt a scheme based on the quantum private channel, for example.

The encryption process can be represented by
\begin{eqnarray}|b\rangle\rightarrow H^{s_1}X^{s_2}|b\rangle,
\end{eqnarray}
where $s_1$ and $s_2$ are a pair of keys, so that the private key with length $k$ can encrypt $\frac{k}{2}$ bits each time.

Similarly, we assume that Eve can distinguish a pseudo-random string whenever its length is more than a sufficiently large $N$. This time she does not need as many as $k\times 2^k\times N$. While each pair of keys has four different possibilities independent from other pairs, 4 ciphertext state can result a right plaintext for testing every possibilities of one pair of keys. Then $4\times N$ ciphertext states encrypted from this pair of keys can certainly result $N$ right plaintexts. Since there are $k/2$ pairs of keys, so $2k\times N$ ciphertext states can fix the whole key. Although these key ciphertexts may be nonadjacent, it would take little more than $N$ ciphertext states to distinguish a pseudo-random string. However, it should be noted that $2k\times N$ is much smaller than $k\times 2^k\times N$.

From this result, we see that the quantum unicity distance of the quantum deterministic private-key encryption scheme may be $O(k)\times N$ which is much smaller than the quantum unicity distance for the probabilistic case, viz., $O(2^k)\times N$.

\section{The Breidbart's attack}
There exists a way \cite{Yan10} to calculate the upper bound of the trace distance of $\rho_0$ and $\rho_1$ defined in Eq.(\ref{rho}). Let
\begin{eqnarray}
\sigma_0=\frac{1}{2^{(k-1)}}\sum_{i}|\varphi_{i_1}\rangle\langle \varphi_{i_1}|\otimes\cdots\otimes|\varphi_{i_k}\rangle\langle\varphi_{i_k}|,
\end{eqnarray}
where $|\varphi_0\rangle=|+\rangle, |\varphi_1\rangle=|1\rangle\}$, $i\in\{0,1\}^k$, and $i_1\oplus i_2\oplus \cdots\oplus i_k=0$.

Similarly,
\begin{eqnarray}
\sigma_0=\frac{1}{2^{(k-1)}}\sum_{i}|\varphi_{i_1}\rangle\langle \varphi_{i_1}|\otimes\cdots\otimes|\varphi_{i_k}\rangle\langle\varphi_{i_k}|,
\end{eqnarray}
where $i_1\oplus i_2\oplus \cdots\oplus i_k=1$.

It can be shown with \cite{Ben96,Iva87,Per88} that
\begin{eqnarray}
D(\sigma_0,\sigma_1)={(\sin\frac{\pi}{4})}^k=(\frac{\sqrt{2}}{2})^k.
\end{eqnarray}

We define a trace-preserving quantum operation $\mathcal{U}$ with operation elements
\begin{eqnarray}
U_{i_1i_2\cdots i_k}=({\frac{\sqrt{2}}{2}})^kH^{i_1}\otimes\cdots\otimes H^{i_k},(i_1,i_2,\ldots,i_k)\in{\{0,1\}}^{k},
\end{eqnarray}
where $H^0$ is unit operator, and $H^1$ is the Hadamard operator:
$$H^1=\frac{\sqrt{2}}{2}\left[\begin{array}{cc}
                             1   & 1\\
                             1   &-1             \end{array}\right].$$

Since
\begin{eqnarray}H^i|+\rangle&=\left\{\begin{array}{ll}
                                                    |+\rangle  &(i=0)\\
                                                    |0\rangle  &(i=1)
                                                    \end{array}\right. \nonumber\\
                H^i|1\rangle&=\left\{\begin{array}{ll}
                                                    |1\rangle  &(i=0)\\
                                                    |-\rangle  &(i=1)
                                                    \end{array}\right.  ,\end{eqnarray}

we have
\begin{eqnarray}
\mathcal{U}(\sigma_0)&=&\sum_{i}U_i\sigma_0U_i^\dag=\rho_0.
\end{eqnarray}

similarly, we have
\begin{eqnarray}
\mathcal{U}(\sigma_1)=\rho_1 .\nonumber
\end{eqnarray}

Then, based on the property of trace-preserving quantum operation,
\begin{eqnarray}\label{max}
D(\rho_0,\rho_1)=D(\mathcal{U}(\sigma_0),\mathcal{U}(\sigma_1))\leq D(\sigma_0,\sigma_1)=(\frac{\sqrt{2}}{2})^k.
\end{eqnarray}

We now use Breidbart's attack to get the lower bound of the trace distance. For $\rho_0^k$ and $\rho_1^k$, when Eve adopts Breidbart's attack, the
difference between probabilities collapsing to $r=0\cdots0$ can be calculated using Eqs.(\ref{rho2}) as:
\begin{eqnarray}
P_{0\cdots0}(\rho_0^k)-P_{0\cdots0}(\rho_1^k)
=\frac{1}{2}(P_{0\cdots0}(\rho_0^{k-1})-P_{0\cdots0}
(\rho_1^{k-1}))(P_0(\rho_0^1)-P_0(\rho_1^1)),\nonumber
\end{eqnarray}
Repeating the iteration, we have
\begin{eqnarray}
P_{0\cdots0}(\rho_0^k)-P_{0\cdots0}(\rho_1^k)
=2\times(\frac{P_0(\rho_0^1)-P_0(\rho_1^1)}{2})^k.
\end{eqnarray}

It is well known that
$$P_0(\rho_0^1)=\cos^2\frac{\pi}{8}, P_0(\rho_1^1)=\sin^2\frac{\pi}{8},$$
so we have
$$P_0(\rho_0^1)-P_0(\rho_1^1)=\frac{\sqrt{2}}{2};$$
similarly, $$P_1(\rho_0^1)-P_1(\rho_1^1)=-\frac{\sqrt{2}}{2}.$$

Hence
$$P_{0\cdots0}(\rho_0^k)-P_{0\cdots0}(\rho_1^k)=2\times(\frac{1}{2}\times\frac{\sqrt{2}}{2})^k.$$

Extending this result to a random $r=r_1\cdots r_k$, the difference is
\begin{eqnarray}
P_{r_1\cdots r_k}(\rho_0^k)-P_{r_1\cdots r_k}(\rho_1^k)=2\times(-1)^{w(r)}(\frac{1}{2}\times\frac{\sqrt{2}}{2})^k.
\end{eqnarray}
where $w(r)$ is the number of 1's in $r$.

Define $\beta_r=P_r(\rho_0), \gamma_r=P_r(\rho_1)$. After Breidbart's attack, the classical trace distance is
$$D(\beta_r,\gamma_r)=\frac{1}{2}\sum_r|\beta_r-\gamma_r|=(\frac{\sqrt{2}}{2})^k.$$

Since $$D(\rho_0,\rho_1)=\max_{\{E_i\}}D(p_r,q_r),$$
this maximization is over all POVMs \{$E_i$\}. We then obtain
\begin{eqnarray}\label{min}
D(\rho_0,\rho_1)\geq (\frac{\sqrt{2}}{2})^k
\end{eqnarray}

From Eqs.(\ref{max}) and (\ref{min}),
\begin{eqnarray}
D(\rho_0,\rho_1)= (\frac{\sqrt{2}}{2})^k
\end{eqnarray}

Note that the POVM of Breidbart's attack results in the maximum Kolmogorov distance over all POVMs. This leads to:
\begin{conjecture}
For a conjugate coding qubit-string, Breidbart's attack may gain most information.
\end{conjecture}

\section{Conclusion}
In this paper, we present a quantum probabilistic encryption algorithm for a private-key encryption scheme based on conjugate coding. We first prove that our scheme can resist attack to the key, invoking the no-signalling postulate. Second, we investigate the scheme's security against plaintext attack, appealing to the concept of information-theoretic indistinguishability of the encryption scheme. Third, we show that, compared with the quantum deterministic private-key encryption scheme, probabilistic encryption greatly increases the unicity distance. Finally, a conjecture is made regarding Breidbart's attack.

\section*{Acknowledgment}
This work was supported by the National Natural Science Foundation of China under Grant No. 61173157.





\bibliographystyle{model1a-num-names}
\bibliography{<your-bib-database>}

\begin{thebibliography}{00}{\footnotesize
\bibitem{Dif76}W. Diffie and M. Hellman, "New directions in cryptography", IEEE Trans. Inform. Theory IT-22, p.644(1976).
\bibitem{Riv78}R. L. Rivest, A. Shamir, and L.A. Adleman, "A Method for Obtaining Digital Signatures and Public-Key Cryptosystems", Commun.ACM {\bf21}, p.210(1978).
\bibitem{God84}S. Goldwasser, S. Micali, "Probabilistic encryption", {\sl Special issue of Journal of Computer and Systems Sciences} {\bf28}(2), p.270(1984).
\bibitem{Yan03}L. Yang, "Quantum Public-Key Cryptosystem Based on Classical NP-Complete Problem", e-print arXiv: quant-ph/0310076.
\bibitem{Nik08}G. M. Nikolopoulos, "Applications of single-qubit rotations in quantum public-key cryptography", Phys. Rev. {\bf A 77}(3), p.032348(2008).
\bibitem{Gao08}F. Gao, Q. Y. Wen, S. J. Qin and F. C. Zhu, "Quantum asymmetric cryptography with symmetric keys", Science in China Series G: Physics Mechanics and Astronomy {\bf52}, p.1925(2008).
\bibitem{Yan1012}L. Yang, M. Liang, B. Li, Lei Hu and D. G. Feng, "Quantum public-key cryptosystems based on induced trapdoor one-way transformations", e-print arXiv: 1012.5249.
\bibitem{Gotte}D. Gottesman, "Quantum pubilic key cryptography with information-theoretic security", unpublished.
\bibitem{Boy03}P. O. Boykin and V. Roychowdhury, "Optimal encryption of quantum bits" Phys. Rev. {\bf A 67}(4), p.42317(2003).
\bibitem{Amb00}A. Ambainis, M. Mosca, A. Tapp and R. Wolf, "Private quantum channel", Proc. 41st FOCS, p.547(2000).
\bibitem{Amb04}A. Ambainis, A. Smith, "Small Pseudo-random Families of Matrices: Derandomizing Approximate Quantum Encryption", Proc. RANDOM, LNCS 3122, Berlin-Heidelberg-NewYork: Springer, p.249(2004).
\bibitem{Wie83}S. Wiesner, "Conjugate coding", SIGACT News {\bf15}, p.78 (1983).
\bibitem{Ben92}C. H. Bennett et al, "Experimental quantum cryptography", J. Cryptology {\bf5}, p.3(1992).
\bibitem{Hut94}B. Huttner and A. Ekirt, "Eavesdropping on quantum-cryptographical systems", J. Mod. Opt {\bf41}, p.2455(1994).
\bibitem{Yan02}L. Yang, L. A. Wu and S. H. Liu, "On the Breidbart eavesdropping problem of the extended BB84 QKD protocol", Acta Phys. Sin. {\bf51}(5), p.961(2002)(in Chinese).
\bibitem{Rab79}M. O. Rabin, "Digitalized Signatures and Public-key Functions as Intractable as Factorization", MIT/LCS/TR-212, MIT Lab, For Computer Science, Cambridge, Mass.(1979).
\bibitem{Gold04}O. Goldreich, {\sl Foudations of Cryptography: Basic Applications}, Publishing House of Electronics Industry, Beijing, 2004.
\bibitem{Yan10}L. Yang, C. Xiang and B. Li, "Qubit-string-based bit commitment protocols with physical security", e-print arXiv: 1011.5099.
\bibitem{sha49}C. E. Shannon, "Communication Theory of Secrecy Systems", Bell Systems Technical Journal {\bf28}, p.656 (1949).
\bibitem{Ben96}C. H. Bennett, T. Mor, J. Smolin, "Parity bit in quantum cryptography", Phys. Rev. {\bf A 54}(4), p.2675 (1996).
\bibitem{Iva87}I. Ivanovic, "How to differentiate between non-orthogonal states", Phys. Lett. {\bf A 123}(6), p.257 (1987).
\bibitem{Per88}A. Peres, "How to differentiate between non-orthogonal states", Phys. Lett. {\bf A 128}(1-2), p.19 (1988).}
\end{thebibliography}







\end{CJK*}
\end{document}